\documentclass[twocolumn,           
               showpacs,            
               preprintnumbers,     
               aps,                 
               prd,          	    
               a4paper,             
               groupedaddress,  
               unsortedaddress,
               nofootinbib,         
               tightenlines,        
               floats               
               ]{revtex4}

\usepackage{graphicx}
\usepackage{amssymb,latexsym}
\usepackage[draft=false]{hyperref}

\newcommand{\ba}{\begin{eqnarray}}
\newcommand{\ea}{\end{eqnarray}}  
\newcommand{\be}{\begin{equation}}
\newcommand{\ee}{\end{equation}}

\begin{document}

\hyphenation{brane-world}  




\title{Observational constraints on braneworld chaotic inflation}
\author{Andrew R.~Liddle and Anthony J.~Smith}
\affiliation{Astronomy Centre, University of Sussex, 
             Brighton BN1 9QJ, United 
Kingdom}
\date{\today} 
\pacs{98.80.Cq \hfill astro-ph/0307017}
\preprint{astro-ph/0307017}


\begin{abstract}
We examine observational constraints on chaotic inflation models in the 
Randall--Sundrum Type II braneworld. If inflation takes place in the high-energy 
regime, the perturbations produced by the quadratic potential are further from 
scale-invariance than in the standard cosmology, in the quartic case more or 
less unchanged, while 
for potentials of greater exponent the trend is reversed. We test these 
predictions against a data compilation including the Wilkinson Microwave 
Anisotropy Probe measurements of 
microwave anisotropies and the 2dF galaxy power spectrum. While in the standard 
cosmology the quartic potential is at the border of what the data allow and all 
higher powers excluded, we find that in the high-energy regime of braneworld 
inflation even the quadratic case is under strong observational pressure. We 
also investigate the 
intermediate regime where the brane tension is comparable to the inflationary 
energy scale, where the deviations from scale-invariance prove to be greater.
\end{abstract}

\maketitle

\section{Introduction}

Braneworld cosmology has opened up a possible new phenomenology for the
cosmology of the early Universe.  Amongst the ideas presently under
investigation, which are nicely reviewed in Ref.~\cite{revs}, are the ekpyrotic 
and cyclic universes \cite{ekpy}, where the Big
Bang may be due to a collision of branes, and various incarnations of braneworld
inflation, where the scalar field may be associated with the distance between
branes \cite{radion}, or may be a bulk field \cite{bulk}, or may live on the
brane \cite{MWBH}.  
 In this paper we explore the simplest and most conservative
scenario, based on the Randall--Sundrum Type II model \cite{RSII} where there is
a single brane upon which the inflaton lives.  In this scenario the detailed
form of the perturbations produced by a given inflationary potential is modified
because the Friedmann equation is modified at high energy, and because the
gravitational wave perturbations are able to penetrate the bulk dimension.

Recently, driven by the announcement of first results from the Wilkinson 
Microwave Anisotropy Probe (WMAP) satellite 
\cite{wmap}, the global cosmological data set has reached a level where it is 
able to significantly constrain inflationary models based on the predicted 
perturbations. Our aim in this short paper is to capitalize on this by obtaining 
observational constraints on some simple braneworld inflation models. We use the 
recently-published constraints of Leach and Liddle \cite{LLnew}, who used a 
compilation of microwave anisotropy data plus the 2dF galaxy power spectrum to 
obtain constraints on the inflationary slow-roll parameters. These results are 
directly applicable also to the braneworld case and so we do not need to repeat 
a data analysis process.

\section{Basic formulae}

We follow the notation set down by Liddle and Taylor \cite{LT}. In the
Randall--Sundrum Type II model \cite{RSII} the Friedmann equation
receives an additional term quadratic in the density \cite{HE}.  The
Hubble parameter $H$ is related to the energy density $\rho$ by
\begin{equation}
\label{Hubble}
H^2 = \frac{8\pi}{3 M_4^2} \, \rho \, \left(1 + \frac{\rho}{2\lambda} \right) 
\,,
\end{equation}
where $M_4$ is the four-dimensional Planck mass 
and $\lambda$ is the brane tension. We have set the four-dimensional 
cosmological constant to zero, and assumed that inflation rapidly makes any dark 
radiation term negligible.
This reduces to the usual Friedmann equation 
for \mbox{$\rho \ll \lambda$}. If the Universe is dominated by a scalar field 
$\phi$ with potential $V(\phi)$, we can use the slow-roll approximation to write 
this as
\begin{equation}
\label{Hubble2}
H^2 \simeq \frac{8\pi}{3 M_4^2} \, V \, \left(1 + \frac{V}{2\lambda} \right) 
\,.
\end{equation}
The 
scalar 
field obeys the usual slow-roll equation
\begin{equation}
3H \dot{\phi} \simeq -V' \,,
\end{equation}
where prime indicates derivative with respect to $\phi$, and dot a 
derivative with respect to time.
The amount of expansion, in terms of $e$-foldings, is given by \cite{MWBH}
\begin{equation}
\label{efolds}
N \simeq - \frac{8\pi}{M_4^2} \int^{\phi_{\rm f}}_{\phi_{\rm i}} \frac{V}{V'}
\left( 1+\frac{V}{2\lambda} \right) d\phi \,,
\end{equation}
where $\phi_{\rm i}$ and $\phi_{\rm f}$ are the values of the scalar
field at the beginning and end of the expansion respectively.

Using the slow-roll approximation as formulated by Maartens et
al.~\cite{MWBH}, the spectra of scalar \cite{MWBH} and tensor
\cite{LMW,HL} perturbations are given by
\begin{eqnarray}
\label{scalamp}
A_{{\rm S}}^2 & = & \frac{4}{25} \, \frac{H^2}{\dot{\phi}^2} \, \left( 
\frac{H}{2\pi} \right)^2 
 \simeq \frac{512 \pi}{75 M_4^6} \, \frac{V^3}{V'^2} \, \left( 1 + 
\frac{V}{2\lambda} \right)^3 \,; \\
A_{{\rm T}}^2 & = & \frac{4}{25\pi} \, \frac{H^2}{M_4^2} \, F^2(H/\mu) 
\label{tensamp} \,,
\end{eqnarray}
where
\begin{eqnarray}
F(x) & = & \left[\sqrt{1+x^2} - x^2 \ln \left( \frac{1}{x} + \sqrt{1 + 
\frac{1}{x^2}} \right) \right]^{-1/2} \,, \nonumber \\
& = & \left[\sqrt{1+x^2} - x^2 \sinh^{-1} \frac{1}{x} \right]^{-1/2} 
\label{functionF}\,,
\end{eqnarray}
and the mass scale $\mu$ is given by
\begin{equation}
\mu = \sqrt{\frac{4\pi}{3}} \, \sqrt{\lambda} \, \frac{1}{M_4} \,.
\end{equation}

The expressions for the spectra are, as always, to be evaluated at Hubble radius 
crossing $k = aH$, and the spectral indices of the scalars and tensors are 
defined as usual by
\begin{equation}
n-1 \equiv \frac{d \ln A_{{\rm S}}^2}{d \ln k} \quad ; \quad n_{{\rm T}} \equiv 
\frac{d \ln A_{{\rm T}}^2}{d \ln k} \,.
\end{equation}
If one defines slow-roll parameters, generalizing the usual ones, by \cite{MWBH}
\begin{eqnarray}
\epsilon_{{\rm B}} & \equiv & \frac{M_4^2}{16\pi} \, \left( 
	\frac{V'}{V} \right)^2 \; \frac{1 + V/\lambda}{\left(1 +
	V/2\lambda \right)^2} \,; \\
\eta_{{\rm B}} & \equiv & \frac{M_4^2}{8\pi} \, \frac{V''}{V} \; 
	\frac{1}{1+V/2\lambda}  \,,
\end{eqnarray}
then the scalar spectral index, in the slow-roll approximation, obeys the usual 
equation
\begin{equation}
n - 1 \simeq -6 \epsilon_{{\rm B}} + 2 \eta_{{\rm B}} \,.
\end{equation}
We define the ratio of tensor to scalar perturbations as
\begin{equation}
R \equiv 16 \, \frac{A_{{\rm T}}^2}{A_{{\rm S}}^2} \,,
\end{equation}
which means our definition of $R$ matches that of Ref.~\cite{LLnew}, with $R 
\simeq 16\epsilon_{{\rm B}}$ in the low-energy limit (note however that that 
paper also defines a slightly different quantity $R_{10}$).

\section{Model predictions}

We restrict our discussion to potentials of the form
\begin{equation}
V = m \phi^\alpha \,,
\end{equation}
where normally $\alpha$ is an even integer, and $m$ is a
constant. This includes the popular quadratic and quartic potentials,
which we will explore in particular detail.

In the standard cosmology, the requirement that the perturbations have the 
observed amplitude fixes the normalization $m$ of the potential. However, in the 
braneworld we additionally have the brane tension $\lambda$. We proceed by 
taking $\lambda$ as a free parameter to be varied, and then adjust the 
normalization of the potential to obtain the correct amplitude of perturbations 
for that $\lambda$. This fixes the inflationary energy scale, whose relation to 
the chosen value of $\lambda$ then determines whether we are in the high- or 
low-energy regime.

With a potential of the above form, setting $\alpha \ge 2$, the
slow-roll parameters are found to satisfy
\begin{equation}
\frac{1}{2}\eta_{\rm B} \le \epsilon_{\rm B} \le 2\eta_{\rm B} \,,
\end{equation}
for any value of $\lambda$. Inflation ends when the slow-roll
conditions, $\epsilon_{\rm B} \ll 1$ and $|\eta_{\rm B}| \ll 1$, are
violated. For ease of computation, we take $\eta_{\rm B} = 1$ to be
the condition for the end of inflation, though it would make no significant 
difference had we adopted the usual $\epsilon_{\rm B} = 1$

The equations simplify significantly in the high- and low-energy
limits, in which we may obtain expressions for $n$ and
$R$ which are independent of $\lambda$ and $m$:
\begin{eqnarray}
n_{{\rm low}} - 1 & = & - \frac{\alpha+2}{2N-1+\alpha} \,; \label{first}\\
n_{{\rm high}} - 1 & = & - \frac{4\alpha+2}{(2+\alpha)N-1+\alpha} \,;\\
R_{{\rm low}} & = & \frac{8\alpha}{2N-1+\alpha} \,;\\
R_{{\rm high}} & = & \frac{24\alpha}{(2+\alpha)N-1+\alpha} \label{last}\,.
\end{eqnarray}
In the limit as $\alpha$ tends to infinity, in the high-energy regime
the scalar spectral index tends to
\begin{equation}
n_{{\rm high}} - 1 = \frac{4}{N+1} \,,
\end{equation}
which corresponds to steep inflation driven by an exponential
potential \cite{CLL}. Table~1 shows some values for particular models.

\begin{table}[b]
\caption[tab1]{\label{tab1} low- and high-energy limits for scalar
spectral index, $n$, and ratio of tensor to scalar perturbations, $R$,
for potentials of the form $V \propto \phi^\alpha$. The end of
inflation is defined by $\eta_{\rm B}=1$ and the number of
$e$-foldings is taken to be 55.}
$$
\begin{array}{c | c c c c}
\alpha & n_\mathrm{low}-1 & n_\mathrm{high}-1 & R_\mathrm{low} &
R_\mathrm{high} \\
\hline
2	&-0.036	&-0.045	&0.144	&0.217\\
4	&-0.053	&-0.054	&0.283	&0.288\\
6	&-0.070	&-0.058	&0.417	&0.324\\
8	&-0.085	&-0.061	&0.547	&0.345
\end{array}
$$
\end{table}

In all the following, we assume that the number of $e$-foldings before
the end of inflation at which observable perturbations are generated
corresponds to $N=55$ \cite{DH,LLe}. The observational values we will
use are defined at about 4 $e$-foldings within the present Hubble
radius \cite{LLnew}. One might have thought that the number 55 ought
to be significantly modified in the case of low $\lambda$ because then
the reheating and radiation eras would at least partly take place in
the high-energy regime, giving a different expansion law. However the
main quantity entering the calculation is the density as a function of
scale factor, rather than of time, which is unchanged, and we find it
is a good approximation to take the number of $e$-foldings as
independent of the brane tension.

In their paper describing the slow-roll formalism for braneworld
models, Maartens et al.~\cite{MWBH} noted that braneworld corrections
tend to drive models towards scale-invariance (i.e.~smaller values of
$R$ and $|n-1|$). While this is true at a given location on the
potential, there is a competing effect that the location of the
potential corresponding to observable perturbations will be closer to
the minimum of the potential, due to the extra friction from the
braneworld term in the Friedmann equation. The above results show that
for small $\alpha$ this latter effect dominates, moving us away from
scale-invariance, whereas for large $\alpha$ it is the former effect
which dominates.

\begin{figure}[t]
\includegraphics[width=\linewidth]{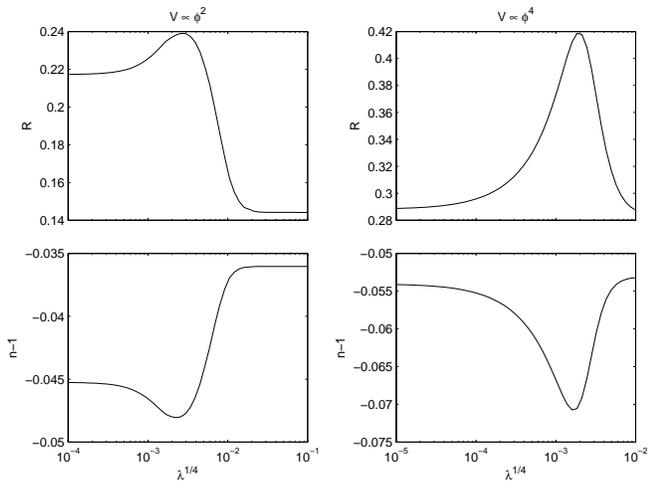}\\
\caption[fig0]{\label{fig0} Theoretical predictions for $n-1$ and $R$
against $\lambda$ for quadratic and quartic potentials. The models are all 
normalized to give the correct perturbation amplitude, and $M_4$ has
been set to be equal to 1.}
\end{figure}

For quadratic and quartic potentials, we have obtained $n$ and $R$ as
functions of the brane tension $\lambda$. This is done firstly by
finding the value of the scalar field at the end of inflation, in
terms of $m$ and $\lambda$, by solving $\eta_{\rm B}=1$ for $\phi$.
Using this, Eq.~(\ref{efolds}) for $N$ can be solved to give
$\phi_{55}(m,\lambda)$, where $\phi_{55}$ is the value of the scalar
field 55 $e$-foldings before the end of inflation. Finally, the COBE
normalization is imposed, in the form $A_{\rm S} = 2 \times 10^{-5}$
\cite{BLW}. Eq.~(\ref{scalamp}) for $A_{\rm S}^2$, which is evaluated at
$\phi = \phi_{55}$, can then be solved numerically to give
$m(\lambda)$. This leaves $\lambda$ as the only free parameter when
determining the predicted perturbations.

Fig.~\ref{fig0} shows the
results for the quadratic and quartic potentials. Large and small values of 
$\lambda$ correspond to the low- and high-energy regimes respectively, with 
asymptotic values for $n$ and $R$ matching the analytic expressions 
Eqs.~(\ref{first})--(\ref{last}). In between there is a continuous curve 
interpolating between the regimes. However note that the interpolation is not 
monotonic; in fact the intermediate regime features greater departures from 
scale-invariance than either of the limits. 

\section{Comparison with observations}

Having made predictions for $n$ and $R$, we are able to compare with 
observational data directly using the recent analysis of Leach and Liddle 
\cite{LLnew}, who used a compilation of microwave anisotropy data including 
WMAP, plus the 2dF galaxy power spectrum, to constrain these parameters. Having 
fixed the number of $e$-foldings of inflation to 55, a given model lives at a 
location in the $n$--$R$ plane, and as $\lambda$ is varied it traces out a 
trajectory in that plane interpolating between low-energy and high-energy 
values. In reality, the points should be somewhat blurred to allow for the 
uncertainty in determining $N$ \cite{LLe}.

\begin{figure}[t]
\includegraphics[width=8cm]{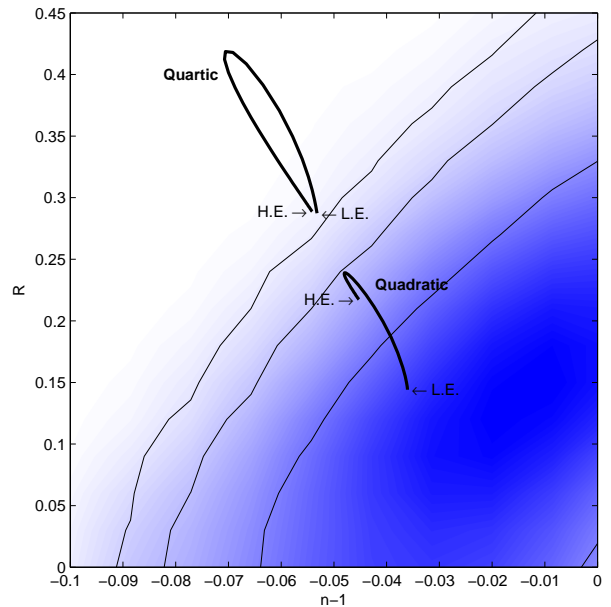}\\
\caption[fig1]{\label{fig1} Theoretical predictions compared to
observational constraints for the quadratic and quartic potentials, as
a function of the brane tension $\lambda$. The low- and high-energy limits are 
shown. The observational contours
are one-, two- and three-sigma confidence levels.}
\end{figure}

In Fig.~\ref{fig1} we show the results for the quadratic and quartic 
potentials.\footnote{This figure differs slightly from Fig.~3 of 
Ref.~\cite{LLnew}; that figure defined $R_{10}$ using the ratio of contributions 
to microwave anisotropies at the tenth multipole, whereas the figures in this 
paper use $R$ defined from the ratio of the power spectra. We thank
Sam Leach for providing the observational constraint data in the present form.} 
The endpoints of the two curves correspond to the low- and high-energy limits 
described in the previous section, and the curves the interpolation between 
them. 
For these potentials, we see that the braneworld moves us further from 
scale-invariance, an effect which can be particularly prominent when the brane 
tension is comparable to the inflationary energy scale.

\begin{figure}[t]
\includegraphics[width=8cm]{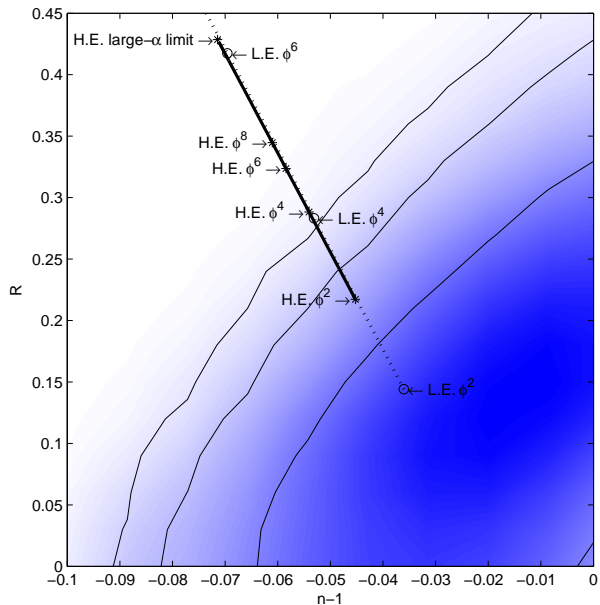}\\
\caption[fig2]{\label{fig2} As Fig.~\ref{fig1}, but now showing the
low-energy limit (dotted line) and high-energy limit (solid line) as a
function of $\alpha$, for $\alpha\ge 2$. We highlight the locations 
corresponding to $\alpha$ being an even integer, and the large-$\alpha$ limit in 
the high-energy case.}
\end{figure}

Fig.~\ref{fig2} shows the low- and high-energy limits as a function of exponent 
$\alpha$, beginning at $\alpha = 2$. We see that the perturbations are much more 
sensitive to $\alpha$ in the low-energy limit than in the high-energy limit, and 
indeed once $\alpha$ exceeds four it is the high-energy limit which is closer to 
scale-invariance. However by this time the models have already moved into the 
observationally excluded region. We therefore conclude that the observational 
upper limit on $\alpha$ is stronger in the braneworld scenario than in the 
standard cosmology, though with present observations the constraint in each case 
lies between $\alpha = 2$ and $4$, precisely where depending on how strict an 
exclusion limit one demands. The quartic potential is however much more strongly 
excluded in the intermediate regime than in either of the limits.

Note that the original steep inflation model with an exponential
potential \cite{CLL} gives the same perturbations as in the high-energy
large-$\alpha$ limit. It is therefore comfortably ruled out by the data,
though steep inflation may still be viable with curvaton reheating
\cite{LU}.

\section{Summary}

In this paper we have computed the perturbation spectra for a set of 
simple braneworld inflation models with monomial potentials, and confronted them 
with the current observational dataset. While na\"{\i}ve expectation might have 
been that the braneworld models gave spectra closer to scale-invariance (as 
preferred by the data), we have found that for small exponents the perturbations 
are further from scale-invariance. Accordingly, observational constraints on the 
exponent are strengthened in the braneworld scenario. While the quadratic 
potential is still allowed at two-sigma for any value of the brane tension, the 
quartic potential is under strong observational pressure, particularly in the 
case where the inflationary energy scale is close to the brane tension.

\begin{acknowledgments}
A.R.L.~was supported in part by the Leverhulme Trust. We thank Sam Leach for 
providing the observational constraints shown in Figs.~\ref{fig1} and 
\ref{fig2}, and Roy Maartens and David Wands for discussions.
\end{acknowledgments}

 

\begin{thebibliography}{}
\bibitem{revs} F. Quevedo, Class. Quant. Grav. {\bf 19}, 5721 (2002),
	{\tt hep-th/0210292}; P. Brax and C. van de Bruck, {\tt hep-th/0303095}.
\bibitem{ekpy} J. Khoury, B. A. Ovrut, P. J. Steinhardt, and N. Turok,
	Phys. Rev. D{\bf 64}, 123522 (2001), {\tt hep-th/0103239}; P. J.
	Steinhardt and N. Turok, Phys. Rev. D{\bf 65} 126003 (2002), {\tt
	hep-th/0111098}.
\bibitem{radion} G. Dvali and S.-H. H. Tye, Phys. Lett. B{\bf 450}, 72 (1999),
	{\tt hep-ph/9812483}; A. Mazumdar and A. P\'erez-Lorenzana, Phys. 
	Lett. B{\bf 508}, 340 (2001), {\tt hep-ph/0102174}.
\bibitem{bulk} H. A. Chamblin and H. S. Reall,  Nucl. Phys. {\bf B562}, 133
	(1999), {\tt hep-th/9903225}; R. N. Mohapatra, A. P\'erez-Lorenzana, 
	and C. A. de S. Pires,
	Phys. Rev. D{\bf 62}, 105030 (2000), {\tt hep-ph/0003089};
	Y. Himemoto and M. Sasaki, Phys. Rev. D{\bf 63}, 044015 (2001),
	{\tt gr-qc/0010035}; Y. Himemoto, T. Tanaka, and M. Sasaki, Phys. 
	Rev. D{\bf 65}, 104020 (2002), {\tt gr-qc/0112027}.
\bibitem{MWBH} R. Maartens, D. Wands, B. A. Bassett, and I. P. C. Heard,
	Phys. Rev. D{\bf 62}, 041301 (2000), {\tt hep-ph/9912464}.
\bibitem{RSII} L. Randall and R. Sundrum, Phys. Rev. Lett. {\bf 83}, 4690 
	(1999), {\tt hep-th/9906064}.
\bibitem{wmap} C. L. Bennett {\it et al.}, {\tt astro-ph/0302207}; D. N. 
	Spergel {\it et al.}, {\tt astro-ph/0302209}; H. V. Peiris 
	{\it et al.}, {\tt astro-ph/0302225}.
\bibitem{LLnew} S. M. Leach and A. R. Liddle, {\tt astro-ph/0306305}.
\bibitem{LT} A. R. Liddle and A. N. Taylor, Phys. Rev. D{\bf 65}, 041301
	(2002), {\tt astro-ph/0109412}.
\bibitem{HE} C. Cs\'aki, M. Graesser, C. Kolda, and J. Terning, Phys. Lett. 
	B{\bf 462}, 34 (1999), {\tt hep-ph/9906513}; J. M. Cline, C. Grojean, 
	and G. Servant, Phys. Rev. Lett. {\bf 83}, 4245 (1999), {\tt
	hep-ph/9906523}; P. Bin\'etruy,
	C. Deffayet, U. Ellwanger, and D. Langlois, Phys. Lett. B{\bf 477}, 
	285 (2000), {\tt hep-th/9910219}; T. Shiromizu, K. I. Maeda, and 
	M. Sasaki, Phys. Rev. D{\bf 62}, 024012 (2000), {\tt gr-qc/9910076}. 
\bibitem{LMW} D. Langlois, R. Maartens, and D. Wands, Phys. Lett. B{\bf 489}, 
	259 (2000), {\tt hep-th/0006007}.
\bibitem{HL} G. Huey and J. E. Lidsey, Phys. Lett. B{\bf 514}, 217 (2001),
	{\tt astro-ph/0104006}.
\bibitem{CLL} E. J. Copeland, A. R. Liddle, and J. E. Lidsey,
	Phys. Rev. D{\bf 64}, 023509 (2001), {\tt astro-ph/0006421}.
\bibitem{DH} S. Dodelson and L. Hui, {\tt astro-ph/0305113}.
\bibitem{LLe} A. R. Liddle and S. M. Leach, {\tt astro-ph/0305263}.
\bibitem{BLW} E. F. Bunn, A. R. Liddle, and M. White, Phys. Rev. D{\bf
	54}, 5917 (1996), {\tt astro-ph/9607038}.
\bibitem{LU} A. R. Liddle and L. A. Urena-Lopez, Phys. Rev. D (to be published),
	{\tt astro-ph/0302054}.
\end{thebibliography}
\end{document}